\documentclass[%
 reprint,
 amsmath,amssymb,
 aps,
]{revtex4-2}
\usepackage{float}
\usepackage{subfigure}
\usepackage{graphicx}
\usepackage{dcolumn}
\usepackage{bm}
\usepackage[colorlinks,
            linkcolor=blue,
            anchorcolor=blue,
            citecolor=blue]{hyperref}
\usepackage{float}
\usepackage{amssymb}
\usepackage{amsmath}
\usepackage{makecell}

\begin{document}
\newtheorem{definition}{Definition} 
\newtheorem{theorem}{Theorem}
\preprint{APS/123-QED}

\title{Dynamical Stability of the Power Law K-essence Dark Energy Model with a New Interaction}
\author{Qile Zhang\textsuperscript{1}}
\email{LOMOSpadeA@outlook.com}
\affiliation{ 1.The Shanghai Key Lab for Astrophysics, 100 Guilin Rd, Shanghai 200234, People’s Republic of China.}

\author{Wei Fang\textsuperscript{1,2}}
\email{Corresponding author:wfang@shnu.edu.cn}
\affiliation{ 1.Department of Physics, Shanghai Normal University, 100 Guilin Rd, Shanghai 200234, People’s Republic of China.
 \\2.The Shanghai Key Lab for Astrophysics, 100 Guilin Rd, Shanghai 200234, People’s Republic of China}

\author{Chenggang Shu\textsuperscript{1}}
\affiliation{ 1.The Shanghai Key Lab for Astrophysics, 100 Guilin Rd, Shanghai 200234, People’s Republic of China.}

\begin{abstract}
We investigate the cosmological evolution of the power law K-essence dark energy (DE) model $F(X)= -\sqrt{X} + X$ with a new interaction $Q = \alpha\rho _m\rho _{\phi }H^{-1}$ in FRWL spacetime. The evolution behavior of dark energy under this interaction is analyzed by using dynamical systems method, and ten critical points are obtained. Among those critical points, a new stable point, which we called Scaling-like dark energy(DE) solution, is very important and interesting. The cosmological meaning of this attractor is different from the Scaling solution and dark energy dominated solution. For some value of model parameters, the universe will evolve to the attractor solution with the dark energy density parameter $\Omega_{\phi}=0.682946$ and the the equation of state $w_{\phi}=-0.99$, which can be in good agreement with the observed data, and alleviate the Coincidence Problem.

\end{abstract}

\maketitle

\section{Introduction}

As is known to all, our universe is nearly flat and experiencing accelerated expansion and contains about $68\%$ dark energy, $27\%$ dark matter, $5\%$ visible matter and negligible radiation\cite{BAHAMONDE20181}. In order to accelerate the expansion of the late universe, the equation of state(EoS) of the late dark energy must be less than $-\frac{1}{3}$. The simplest dark energy candidate is a very small positive cosmological constant($\Lambda$) which EoS is constant $-1$. But it's hard to understand why the cosmological constant is about 120 orders of magnitude smaller than the theoretical expectation of Planck's energy density, this is the Cosmological Constant Problem. Another problem related to dark energy is the Coincidence Problem, that is, why the density of dark energy and the density of matter are now on the same order of magnitude. Because the evolution of the two is so different with the time of the universe, in order to make them in the same order of magnitude today, we must fine tune the dark energy density to an unacceptable level in the early universe.

$Planck$ 2018\cite{osti_1676387} shows that $w=-1.028 \pm 0.031$ ($68\%$, $Planck$ TT, TE, EE+lowE + lensing + SNe + BAO), that is, we can't exclude the current equation of state of dark energy passing through $- 1$\cite{PhysRevD.68.023522}. The recent Hubble constant tension has also troubled many scientists, Planck and the standard cosmological model give the prediction of Hubble constant $H_{0}=67.4 \pm 0.6 \mathrm{~km} / \mathrm{s} / \mathrm{Mpc}$, different from $H_{0}=74.03 \pm 1.42 \mathrm{~km} / \mathrm{s} / \mathrm{Mpc}$ obtained from SNe Ia Supernova. Rencently the early dark energy model (EDE) has attracted people's attention, that is, a scalar field with up to approximately $10\%$ relative energy density at a critical redshift $z_{c}$ about 3500 seems able to bring the high and low redshift measurements of $H_{0}$ into agreement which can alleviate the Hubble Tension\cite{PhysRevD.103.043518,PhysRevD.102.043507,PhysRevD.102.103502,PhysRevD.101.043533,jedamzik2021reducing,smith2020early}.

K-essence is characterized by a scalar field with a non-canonical kinetic energy, such as Low-energy effective string theory with derivative terms higher-order than $X$, Ghost condensate model,Tachyon field, Dirac-Born-Infeld(DBI) theories. Usually K-essence models has the Lagrangian density with the form $\mathcal{L}= F(X)V(\phi)$\cite{article4, PhysRevD.89.123514,2014JPhCS.485a2017D} which $F(X)= - X + X^2$. in this paper we adopt $F(X)=-\sqrt{X} + X$\cite{Yang_2014} , which has many interesting properties worth studying.

This paper is organized as follows: in the second section, we simply introduce the dynamical systems w and its stability analysis. In the third section, we consider the dynamics of the k-essence scalar field with the new interaction $Q=\alpha\rho _m\rho _{\phi }H^{-1}$ \cite{2015IJMPD..2430007B,Perez_2014}. Finally, we will present the conclusion and the discussion of our model.

\section{Dynamical systems and its stability analysis}

\subsection{What are dynamical systems}

 In general, we can think of a dynamical system as any abstract system consisting of
\\\textbf{1.} a space (state space or phase space), and
\\\textbf{2.} a mathematical rule describing the evolution of points in that space.

\subsection{Autonomous equation}
First, we consider two dimensional system:
\begin{equation}
\begin{array}{l}
\dot{x}=f(x, y) \\
\dot{y}=g(x, y)
\end{array}
\end{equation}

In general, the first derivative of $x$ and $y$ with respect to $t$ are $f (x, y, t)$ and $g (x, y, t)$, but if $f (x, y, t)$ and $g (x, y, t)$ are not explicit time $t$ which means $\dot{x}=f(x, y)$ and $\dot{y}=g(x, y)$, $\dot{} \equiv \frac{\mathrm{d}}{\mathrm{d} t}$, the two-dimensional dynamic equation becomes special, we call it autonomous equation.
\\\textbf{Definition :} The autonomous equation $\dot{x}=f(x, y)$ and $\dot{y}=g(x, y)$ are said to have a \textbf{critical (or fixed or equilibrium) point} at $x=x_0$ and $y=y_0$ if and only if $f(x_0, y_0)=0$ and $g(x_0, y_0)=0$.

However, above \textbf{Definition} could not guarantee that the dynamical system remain in this (steady) state indefinitely. One needs to clarify whether or not the system can in fact attain such a state and whether or not this state is stable with respect to small perturbations.
This naturally leads to the question of stability of a critical point or fixed point. In simple words a fixed point $(x_0, y_0)$ of the system (1) is called \textbf{Stable Point} if all solutions $(x(t), y(t))$ starting near $(x_0, y_0)$ stay close to it.
\subsection{stability}
The stability of the fixed point can be obtained by solving the Jacobian matrix of the system:
\begin{equation}
\begin{array}{l}
\dot{x}=f(x, y) \\
\dot{y}=g(x, y)
\end{array}
\end{equation}
Its Jacobian matrix is $J$
\begin{equation}
\left[\begin{array}{ll}
\frac{\partial f}{\partial x} & \frac{\partial f}{\partial y} \\
\frac{\partial g}{\partial x} & \frac{\partial g}{\partial y}
\end{array}\right]_{\left(x_0, y_0\right)}
\end{equation}
After calculating the fixed points of the system, each point is substituted into the Jacobian matrix
\\\indent \textbf{(i)} \textbf{Node point} : If all eigenvalues of Jacobian matrix are negative(positive), we called the critical point stable(unstable) node.
\\\indent \textbf{(ii)} \textbf{Spiral point} : If the determinant is negative and the real parts of all eigenvalues are negative(positive), the critical point is stable(unstable).
\\\indent 
\\\indent \textbf{(iii)} \textbf{Saddle Point} : If the sign of the real part of two eigenvalues is opposite. A saddle point is a point that behaves as an attractor for some trajectories and a repellor for others.

For the dynamic matrix of two-dimensional system, the eigenvalue equation is 

\begin{equation}
\lambda^{2}+p \lambda + q = 0
\end{equation}
where
\begin{equation}
p \equiv -(\frac{\partial P}{\partial x}+\frac{\partial Q}{\partial y})_{\left(x_0, y_0\right)} = -(\lambda_{1}+\lambda_{2}) = -Tr(J)
\end{equation}
\begin{equation}
q \equiv (\frac{\partial P}{\partial x} \frac{\partial Q}{\partial y}-\frac{\partial P}{\partial y} \frac{\partial Q}{\partial x})_{\left(x_0, y_0\right)} = \lambda_{1} \lambda_{2} = Det(J)	
\end{equation}
 
If we use $Tr(J)$ and $Det(J)$ to judge whether this point is a stable node, an unstable node or a saddle node, we can obtain that, if this point is stable node, $Det(J)$ should be positive and $Tr(J)$ should be negative. if this point is unstable node, $Det(J)$ should be positive and $Tr(J)$ should be also positive. if this point is saddle node, $Det(J)$ should be negative and $Tr(J)$ can be positive or negative.
\begin{table}[htbp]
\begin{ruledtabular}
\begin{tabular}{ccc}
\textrm{\textbf{Stability}}&
\textrm{\textbf{$Tr(J)$}}&
\textrm{\textbf{$Det(J)$}}\\
\colrule
stable node & $-$ & $+$\\
unstable node & $+$ & $+$\\
saddle point& $+/-$ & $-$\\
\end{tabular}
\end{ruledtabular}
\end{table}

In fact, we're more interested in attractors, so in this paper, we will not find out the specific value of each Jacobian matrix eigenvalue, but if the trace of the matrix is negative($Tr(J)<0$) and the determinant is positive($Det(J)>0$), we can judge that the point is a stable point\cite{PhysRevD.57.4686,universe6120244}.
\section{Model}
\subsection{K-essence}
Scalar fields with non-canonical kinetic terms often appear in particle physics. In general the action for such theories can be expressed as
\begin{equation}
S=\int \mathrm{d}^{4} x \sqrt{-g}\left[\frac{1}{2 \kappa^{2}} R+\mathcal{L}(\phi, X)\right]+S_{M}	
\end{equation}

where $\mathcal{L}(\phi, X)$ is a function in terms of a scalar field $\phi$ and its kinetic energy $X\equiv -\frac{1}{2}(\nabla \phi)^{2}$, and $S_{M}$ is a matter action. Even in the absence of the field potential $V(\phi)$ it is possible to realize the cosmic acceleration due to the kinetic energy $X$\cite{Armend_riz_Pic_n_1999}. The application of these theories to dark energy was carried out by Chiba\cite{Chiba:1999ka}. In Ref\cite{ArmendarizPicon:2000dh}, this was extended to more general cases and the models based on the action (7) were named “K-essence”. 

There are many forms of K-essence model, such as $\mathcal{L}=X G(X / V(\phi))$\cite{article2, article3, Chiba:1999ka}, $\mathcal{L}= F(X)V(\phi)$\cite{article4, PhysRevD.89.123514}, $\mathcal{L}= F(X)-V(\phi)$\cite{2014JPhCS.485a2017D}. This paper will adopt the following form
\begin{equation}
\mathcal{L}= F(X)V(\phi)	
\end{equation}
Unlike Quintessence\cite{PhysRevLett.80.1582, PhysRevD.59.123504, PhysRevLett.82.896}, the form of potential energy $V(\phi)$ determines the universe evolution, the universe history depends on $V(\phi)$ and $F(X)$.
\subsection{The Power Law K-Essence Dark Energy Model and Its Stability Analysis}
We consider a kind of Lagrangian quantity in the form of
\begin{equation}
\mathcal{L}(\phi, X)=(-\sqrt{X} + X)V(\phi)
\end{equation}
It was proposed by Bose and Majumdar\cite{PhysRevD.80.103508}. For relevant properties, please refer to Yang\cite{Yang_2014}. In Yang's paper, there is no interaction between dark energy and matter. An interaction between dark energy and matter with the form $Q=\alpha H \rho _m$ is considered\cite{universe6120244} , In fact, the interaction  $Q=\beta H \rho _\phi$ can also be considered. However, in this paper we will consider a new interaction $Q=\alpha\rho _m\rho _{\phi }H^{-1}$, which is related to both dark energy and matter term. We will work with a flat, homogeneous, and isotropic FRWL spacetime having signature $(-,+,+,+)$, and in units $c$ = $8\pi G$ = 1. The energy density of scalar field, the EoS and the corresponding sound velocity are
\begin{equation}
\rho_{\phi} =V(\phi)\left[2 X F_{X}-F\right]=X V
\end{equation}
\begin{equation}
w_{\phi} =\frac{F}{2 X F_{X}-F}=\frac{X-\sqrt{X}}{X}
\end{equation}
\begin{equation}
c_{\mathrm{s}}^{2} =\frac{\partial p / \partial X}{\partial \rho / \partial X}=\frac{F_{X}}{F_{X}+2 X F_{X X}}=1-\frac{1}{2 \sqrt{X}}
\end{equation}
which
\begin{equation}
F_{X} \equiv \frac{d F}{d X}\quad , \quad F_{X X} \equiv \frac{d^{2} F}{d X^{2}}
\end{equation}
the auxiliary variables are defined as
\begin{equation}
 x={\dot{\phi}}\quad 
   \quad 
 y=\frac{\sqrt{V}}{\sqrt{3} H}
\end{equation}
Then we get
\begin{equation}
H=\frac{\sqrt{V}}{\sqrt{3} y},\quad
\rho _{\phi }=\frac{x^2}{2}V,\quad
\rho _m=(\frac{1}{y^2}-\frac{x^2}{2})V
\end{equation}
and finally we have 
\begin{equation}
  w_{\phi}=1-\frac{\sqrt{2}}{|{x}|}
  \end{equation}
 
\begin{equation}
  \Omega_{\phi} = \frac{1}{2}x^2y^2
  \end{equation}
   
\begin{equation}
  c^2_{s}=1-\frac{\sqrt{2}}{2|{x}|}
  \end{equation}
 
 Obviously, unlike the $\mathcal{L}(\phi, X)=(-X^{2} + X)V(\phi)$, we know that if $ \dot{\phi} >0 $, then $\sqrt{X} = \frac{\sqrt{2}}{2}x $, if $ \dot{\phi} <0 $, then $\sqrt{X} = -\frac{\sqrt{2}}{2}x $.

The friedmann equations take the form
\begin{equation}
H^{2}=\frac{1}{3}\left(\rho_{\mathrm{m}}+\rho_{\phi}\right)
\end{equation}
\begin{equation}
\dot{H}=-\frac{1}{2}\left(\rho_{\mathrm{m}}+\rho_{\phi}+p_{\phi}\right)
\end{equation}
The equation of motion for the k-essence field is given by
\begin{equation}
\left(F_{X}+2 X F_{X X}\right) \ddot{\phi}+3 H F_{X} \dot{\phi}+\left(2 X F_{X}-F\right) \frac{V_{\phi}}{V}=0
\end{equation}

With the interaction $Q$ between dark energy and matter term, $\rho_{\mathrm{m}}$ and $\rho_{\phi} $ do not satisfy conservation equations  separately, and the following two equations are satisfied:

\begin{equation}
\begin{aligned}
\dot{\rho}_{\phi}+3 H\left(\rho_{\phi}+p_{\phi}\right) &=Q \\
\dot{\rho}_{m}+3 H\left(\rho_{m}+p_{m}\right) &=-Q
\end{aligned}
\end{equation}
here $Q=\alpha\rho _m\rho _{\phi }H^{-1}$. There is no interaction between dark energy and matter term when $Q=0$.  If $Q>0$, that means there is a transfer of energy from dark matter to dark energy. $Q<0$, there is a transfer of energy from dark energy to dark matter. Some restrictions\cite{Garriga:1999vw}:
\\\indent \textbf{1}.$x, y \geq 0$
\\\indent \textbf{2}.$0\le\Omega_{\phi} = \frac{1}{2}x^2y^2\le1$
\\\indent \textbf{3}.$0\le c^2_{s}=1-\frac{\sqrt{2}}{2}{x}\le1$
\\\indent \textbf{4}.$w_{\phi}=1-\frac{\sqrt{2}}{x}\le -\frac{1}{3}$
  
 We can get the following dynamical system from Eq.(14) and Eq.(21):
\begin{align}
x^{\prime} &= \frac{\sqrt{3}}{2}  \lambda  x^{2}  y-3  x+\frac{3 \sqrt{2}}{2}+\frac{\sqrt{3}  y Q}{x 	 V^{\frac{3}{2}}}\\
y^{\prime} &= \frac{1}{4} y(6+x y(3(-\sqrt{2}+x) y-2 \sqrt{3} \lambda))
\end{align} 
 where prime $'$ means the derivative with respect to $ln a$. Obviously,  our new interaction $ Q=\alpha\rho _m\rho _{\phi }H^{-1}$ can make the dynamical system be closeness and an autonomous dynamical system. Potential parameter $\lambda  \equiv -\frac{V_{, \phi}}{V^\frac{3}{2}}$. In this paper we chose the form of potential as $V(\phi) \propto  \phi^{-2}$\cite{PhysRevD.89.123514} and then $\lambda=constant$.

\subsection{The Analysis of Stability for This Dark Energy Model with Interaction $Q=\alpha\rho _m\rho _{\phi }H^{-1}$}
\subsubsection{$x>0$}
\begin{align}
x^{\prime} &= \frac{1}{4}[-3 \alpha  x^3 y^2+2 \sqrt{3} \lambda  x^2 y+6 (\alpha -2) x+6 \sqrt{2}]\\
y^{\prime} &= \frac{1}{4} y(6+x y(3(-\sqrt{2}+x) y-2 \sqrt{3} \lambda))
\end{align}
\begin{table*}
\caption{Points.}
\begin{ruledtabular}
\begin{tabular}{cccccc}
\hline
  Point & $\Omega_{\phi}$ & $w_{\phi}$ & $c^2_{s}$ & Existence and Stability & \quad \\ \hline
  $x>0$  \\ \hline
 $P_1$ & $\frac{\left(\sqrt{(\alpha -1)^2 \left(\lambda ^2-6 \alpha \right)}+(\alpha -1) \lambda \right)^2}{6 (\alpha -1)^2 \alpha ^2}$ & $\alpha $ &$\frac{\alpha +1}{2}$&$-1<\alpha <-\frac{1}{3}\land \lambda >\sqrt{\frac{3}{2}} \sqrt{\alpha ^2+2 \alpha +1}$\\
 $P_2$ & $\frac{\left(\sqrt{(\alpha -1)^2 \left(\lambda ^2-6 \alpha \right)}+(1-\alpha)\lambda \right)^2}{6 (\alpha -1)^2 \alpha ^2}$ &  $\alpha $ &$\frac{\alpha +1}{2}$&None \\
$P_3$ & $1$ & $\sqrt{\frac{2}{3}} \lambda -1$ & $\frac{\lambda }{\sqrt{6}}$ & \makecell[c]{$\left(-1<\alpha \leq -\frac{1}{3}\land 0<\lambda <\frac{1}{4} \left(\sqrt{6} \alpha +3 \sqrt{6}\right)-\frac{1}{4} \sqrt{6 \alpha ^2-12 \alpha +6}\right)$or \\$ \left(\alpha >-\frac{1}{3}\land 0<\lambda <\sqrt{\frac{2}{3}}\right)$} \\
$P_4$ & $1$ & $-\sqrt{\frac{2}{3}} \lambda -1$ & $-\frac{\lambda }{\sqrt{6}}$ &None \\
$P_5$ & $0$ & $\alpha -1$ & $\frac{\alpha }{2}$&None \\ \hline
 
  $x<0$  \\ \hline
 $P_6$ & $\frac{\left(\sqrt{(\alpha -1)^2 \left(\lambda ^2-6 \alpha \right)}+(1-\alpha)\lambda \right)^2}{6 (\alpha -1)^2 \alpha ^2}$ & $\alpha $ &$\frac{\alpha +1}{2}$&$-1<\alpha <-\frac{1}{3}\land \lambda <-\sqrt{\frac{3}{2}} \sqrt{\alpha ^2+2 \alpha +1}$\\
 $P_7$ & $\frac{\left(\sqrt{(\alpha -1)^2 \left(\lambda ^2-6 \alpha \right)}+(\alpha -1) \lambda \right)^2}{6 (\alpha -1)^2 \alpha ^2}$ &  $\alpha $ &$\frac{\alpha +1}{2}$ &None \\
$P_8$ & $1$ & $-\sqrt{\frac{2}{3}} \lambda -1$ & $-\frac{\lambda }{\sqrt{6}}$ &None \\
$P_9$ & $1$ & $\sqrt{\frac{2}{3}} \lambda -1$ & $\frac{\lambda }{\sqrt{6}}$ &\makecell[c]{$\left(-1<\alpha \leq -\frac{1}{3}\land \frac{1}{4} \sqrt{6 \alpha ^2-12 \alpha +6}+\frac{1}{4} \left(\sqrt{6} (-\alpha )-3 \sqrt{6}\right)<\lambda <0\right)$or\\$ \left(\alpha >-\frac{1}{3}\land -\sqrt{\frac{2}{3}}<\lambda <0\right)$} \\
$P_{10}$ & $0$ & $\alpha -1$ & $\frac{\alpha }{2}$ &None \\ \hline
\end{tabular}
\end{ruledtabular}
\end{table*}

\begin{figure}[H]
\centering
\subfigure[The value ranges for parameters $\lambda$ and $\alpha$ to make the critical point $P_1$ stable, which is also constrained by other cosmological quantities, when $x >0$.]{\includegraphics[width=0.5\textwidth]{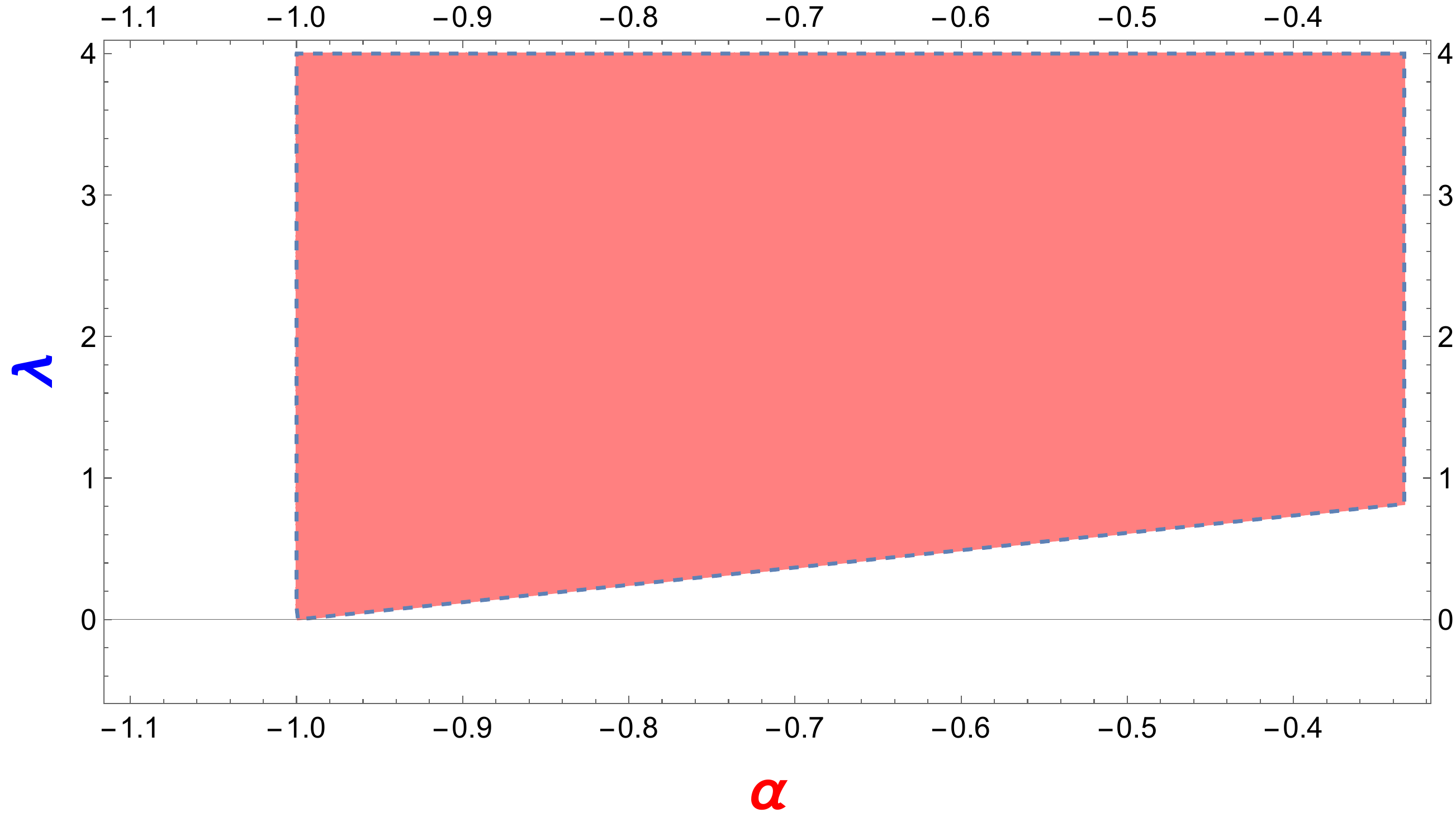}}

\subfigure[The phase plane for $\left\{\alpha \rightarrow-\frac{99}{100}, \lambda \rightarrow \frac{12}{25}\right\}$ around the attractor $P_1$ = $(0.71066, 1.64455)$ when $x > 0$. the red point is the final stable point.]{\includegraphics[width=0.5\textwidth]{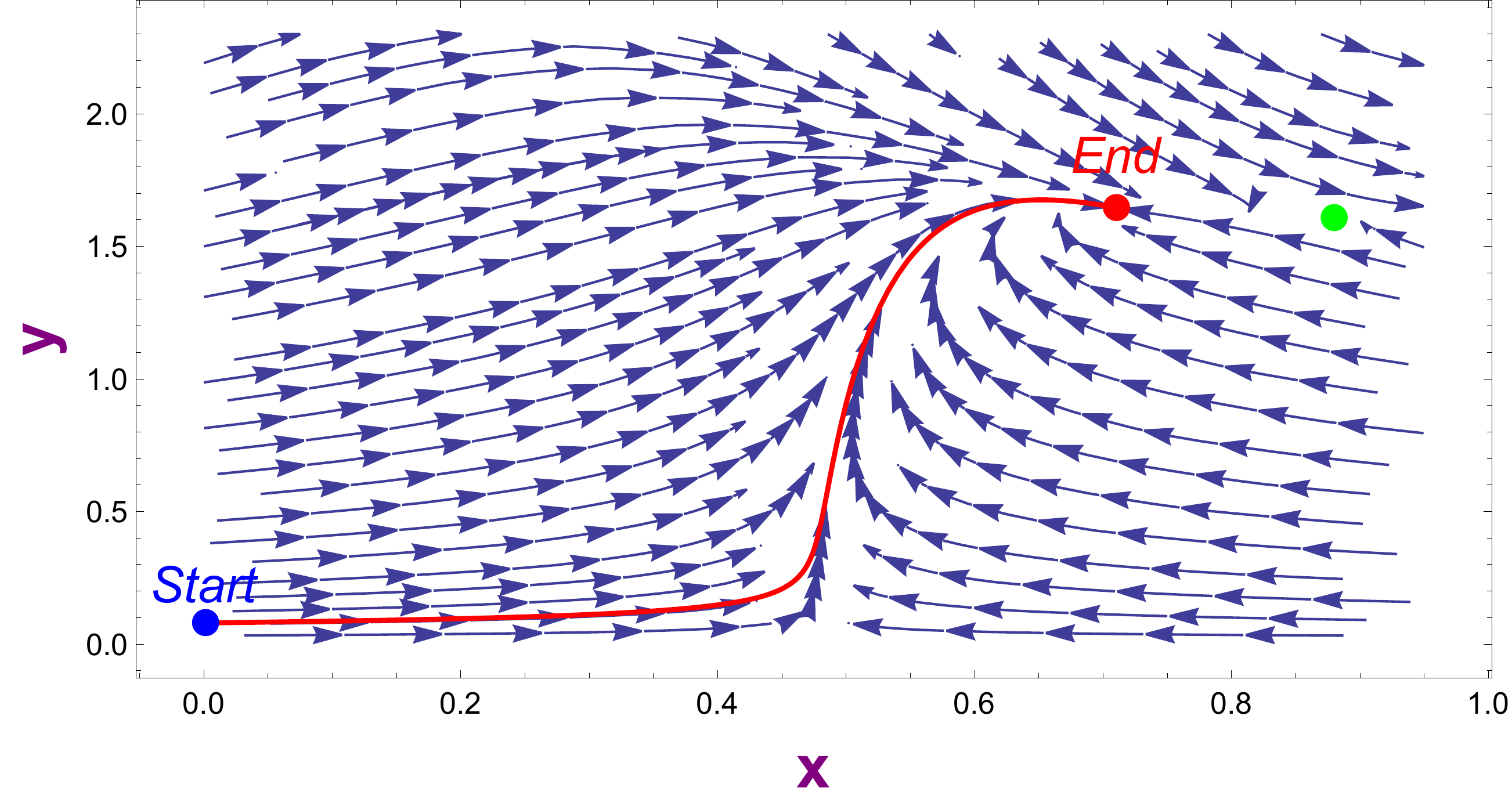}}
\caption{\label{fig:1} Parameter space(a) and phase plane(b) for critical point $P_1$}
\end{figure}
\begin{figure}[h]
\centering
\subfigure{\includegraphics[width=0.5\textwidth]{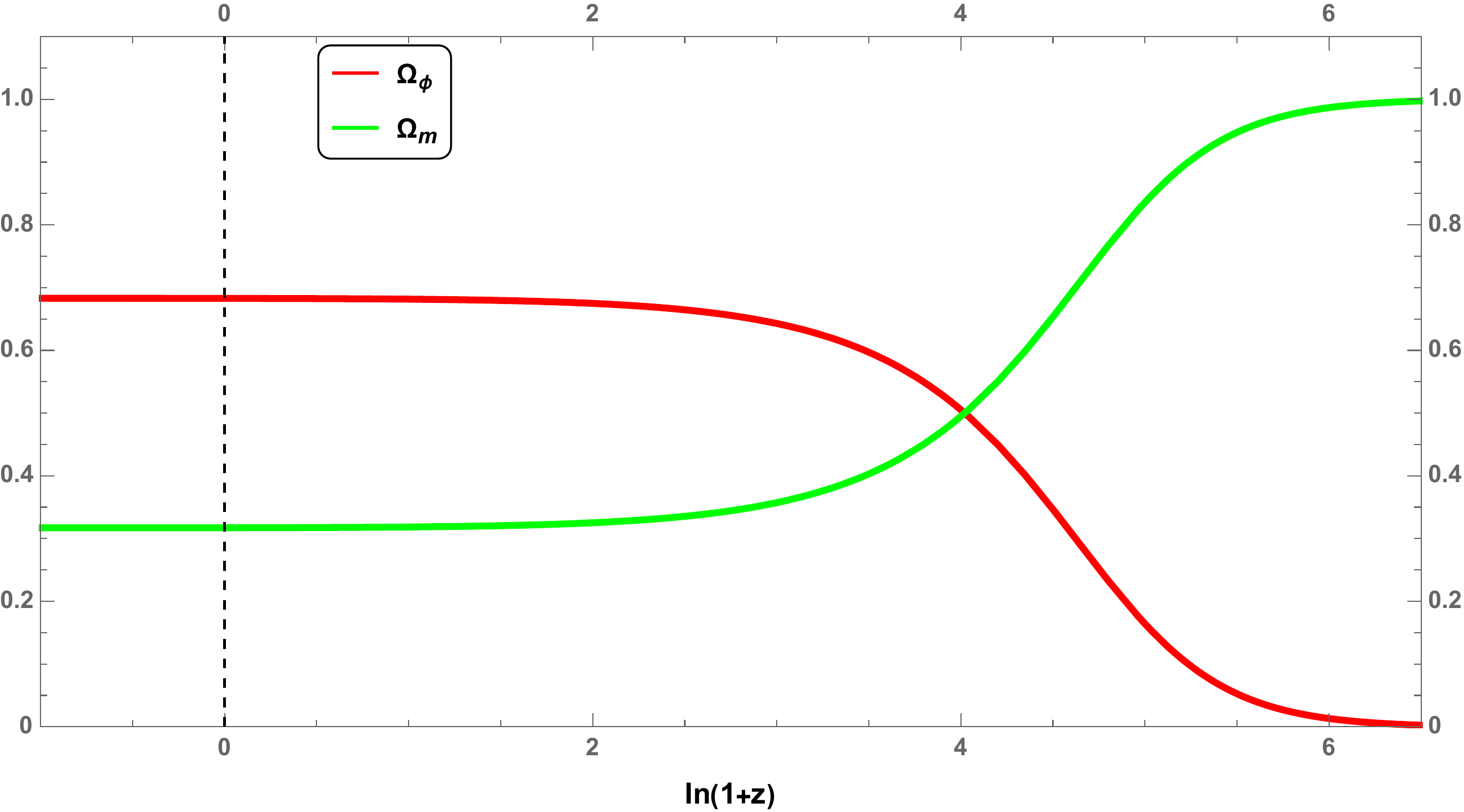}}
\subfigure{\includegraphics[width=0.5\textwidth]{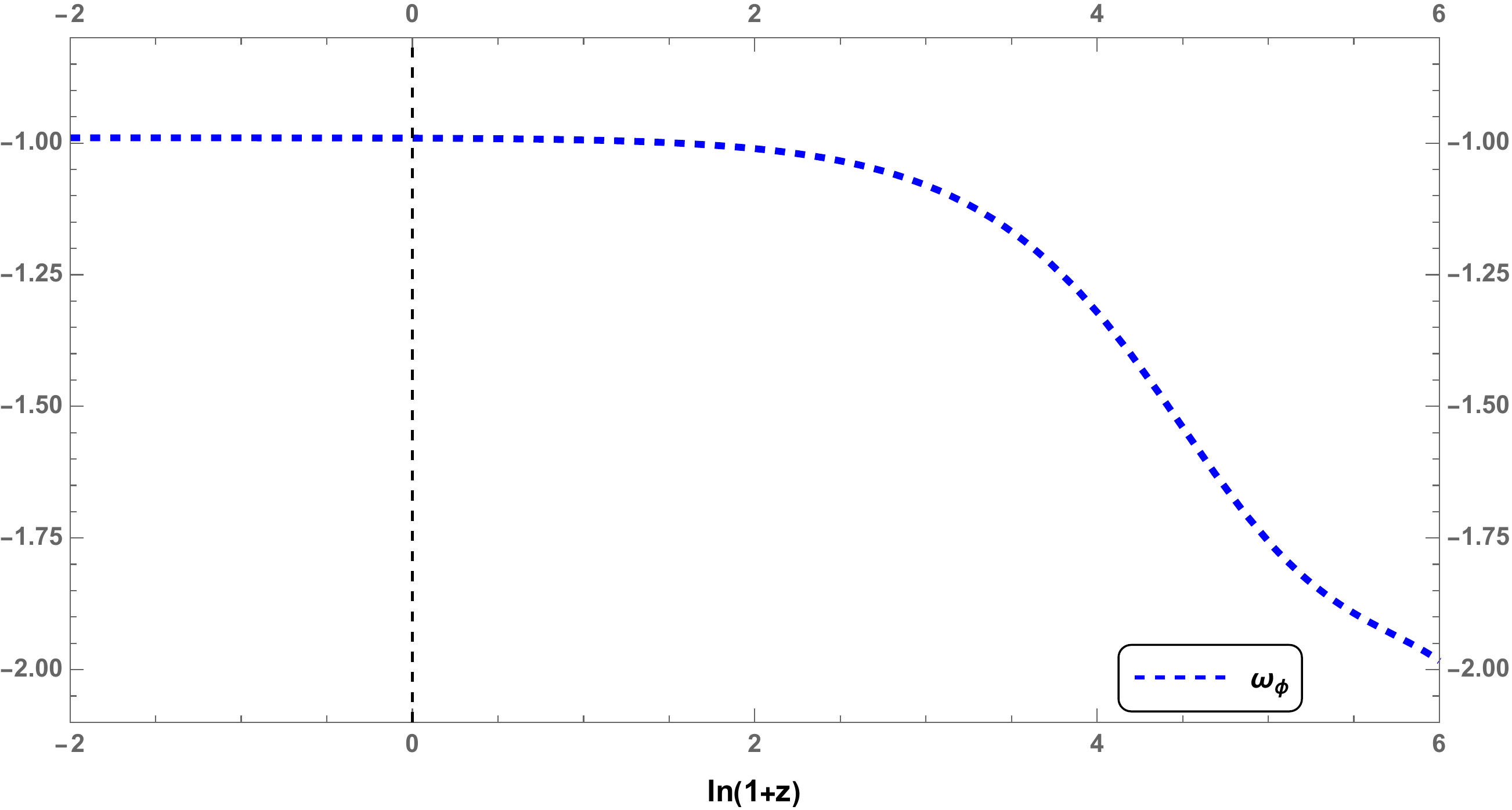}}
\caption{\label{fig:2} $P_1$:
The evolution of $\Omega_{\phi}$, $\Omega_{m}$ and $w_{\phi}$ for $\left\{\alpha \rightarrow-\frac{99}{100}, \lambda \rightarrow \frac{12}{25}\right\}$ with the initial conditions $x = 0.00065$ and $y = 0.08$ when $ln (1+z) = -7$.}
\end{figure}
 We only consider the stable critical points with the possibility of accelerating expansion phase.
 
$P_1$: Point $P_1$ is a very interesting point which does not exist in the power law k-essence model with no interaction. We define $P_1$ as a Scaling-like Dark energy(Scaling-like DE) solution, different from the $Scaling$ solution\cite{Wetterich_1988}. To be a stable point with cosmological meaning, it should satisfy the condition of $ Det >0$, $ Tr<0 $ and the four constraints mentioned above. We finally get the feasible parameter regions which satisfy these conditions as shown in Fig\ref{fig:1}, select the feasible parameters and draw the  phase plane, as shown in Fig\ref{fig:1}. We can see  from the trajectory that $P_1$ is really a stable point, and the trajectories coming from different directions will be attracted to the red point $P_1$. The universe will eventually evolve to the state of the red point $P_1$. The green point near point $P_1$ is a saddle point(In fact, the green dot is the point $P_3$ under the values of $\left\{\alpha \rightarrow-\frac{99}{100}, \lambda \rightarrow \frac{12}{25}\right\}$). Obviously our universe will not reach this point($P_3$). We calculate that the coordinates of the attractor are $(0.71066, 1.64455)$, and the corresponding $\Omega_{\phi}$ and $w_{\phi}$ are $0.682946$ and $-0.99$. The physical meaning of this point is that at the moment when background matter transforms into dark energy, the universe will eventually neither be dominated by dark matter nor dark energy, but due to the existence of negative pressure dark energy, it can accelerate the expansion of the late universe. This is the reason we called point $P_1$ Scaling-like Dark energy(Scaling-like DE) solution.

According to the initial conditions $x = 0.00065$ and $y = 0.08$ when $ln (1+z) = 7(z\approx1100)$, we calculate the present value of $(x, y) $ as $(0.710428, 1.64487)$.  As shown in Fig \ref{fig:2}, we find that in the early stage, $w_{\phi}$ is less than $-1$ in the early stage, When the redshift at $ln (1+z)$ is less than 5$(z\approx150)$, dark matter begins to gradually transform into dark energy, $w_{\phi}$ can cross $-1$. Substituting the initial conditions we choose, we get that $\Omega_{\phi}$ is now $0.682767$ and $w_{\phi}$ is $-0.99065$. The huge advantage for $P_1$ is that it can give the solution to the famous coincidence problem since the current state of our universe may be in the basin of attraction very near to point $P_1$. That means we do not need the fine tune.

$P_2$: Considering the four constraints mentioned above, the parameter ranges of the four conditions have no intersection which means this point does not meet the late cosmic acceleration conditions we require, so we ignore it.

\begin{figure}[h]
\centering
\subfigure[The value ranges for parameters $\lambda$ and $\alpha$ to make the critical point $P_3$ stable, which is also constrained by other cosmological quantities, when $x >0$.]{\includegraphics[width=0.5\textwidth]{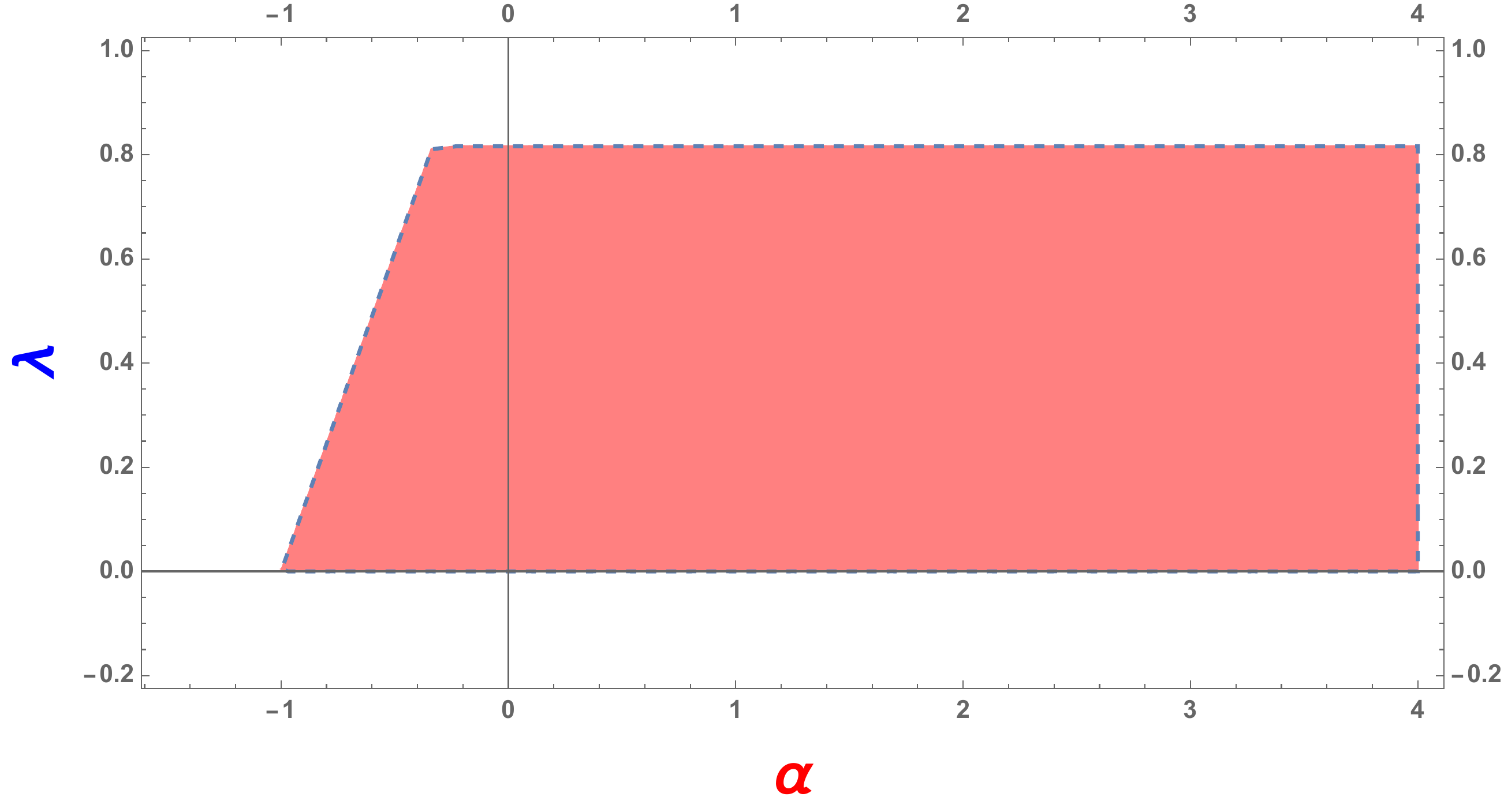}}

\subfigure[The phase plane for $\left\{\alpha \rightarrow\frac{1}{20}, \lambda \rightarrow \frac{1}{100}\right\}$ around the attractor $P_3$ = $(0.710005, 1.99184)$ when $x > 0$.]{\includegraphics[width=0.5\textwidth]{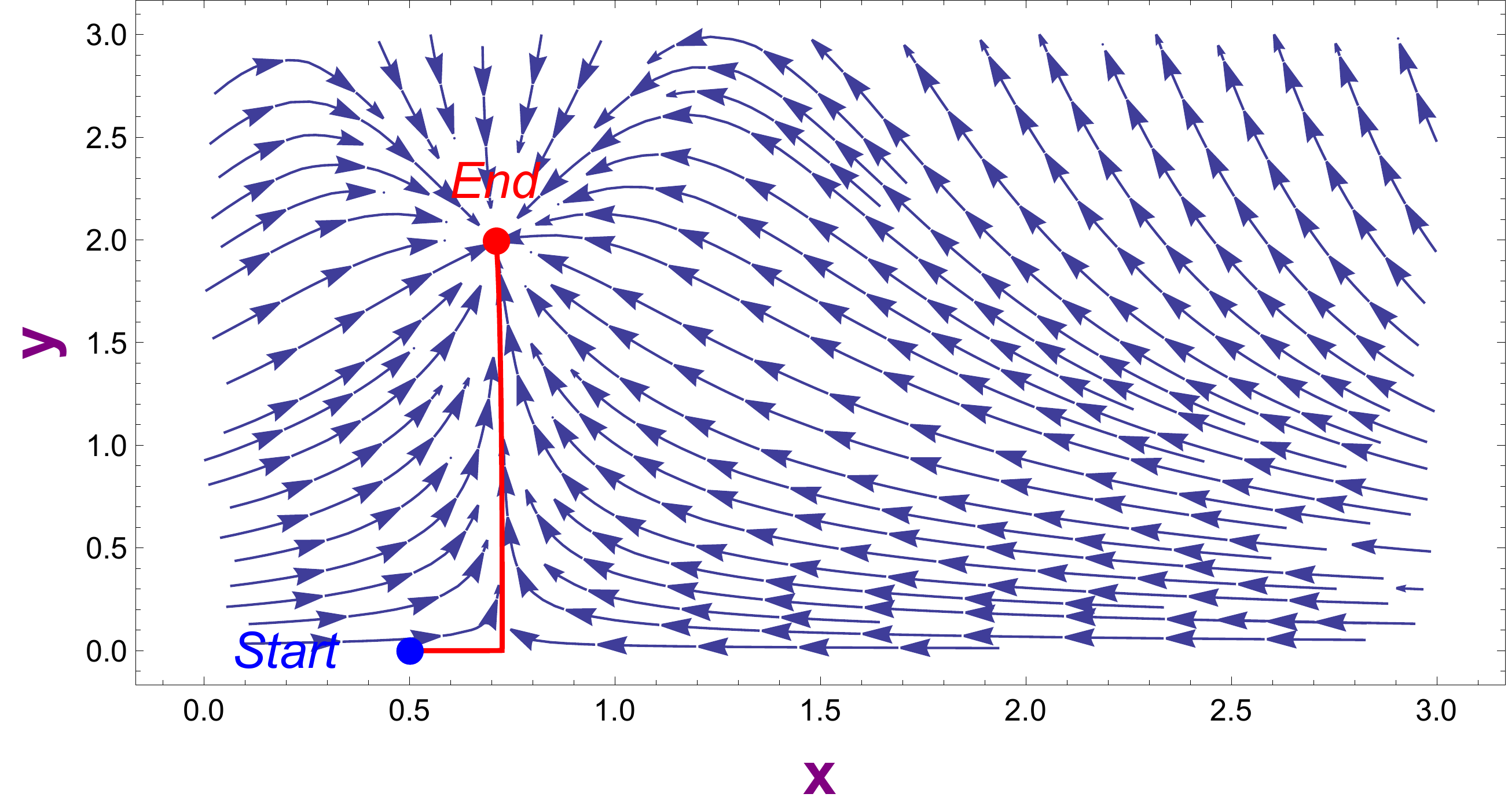}}
\subfigure[The phase plane for $\left\{\alpha \rightarrow-\frac{1}{2}, \lambda \rightarrow \frac{1}{100}\right\}$ around the attractor $P_3$ = $(0.710005, 1.99184)$ when $x > 0$]{\includegraphics[width=0.5\textwidth]{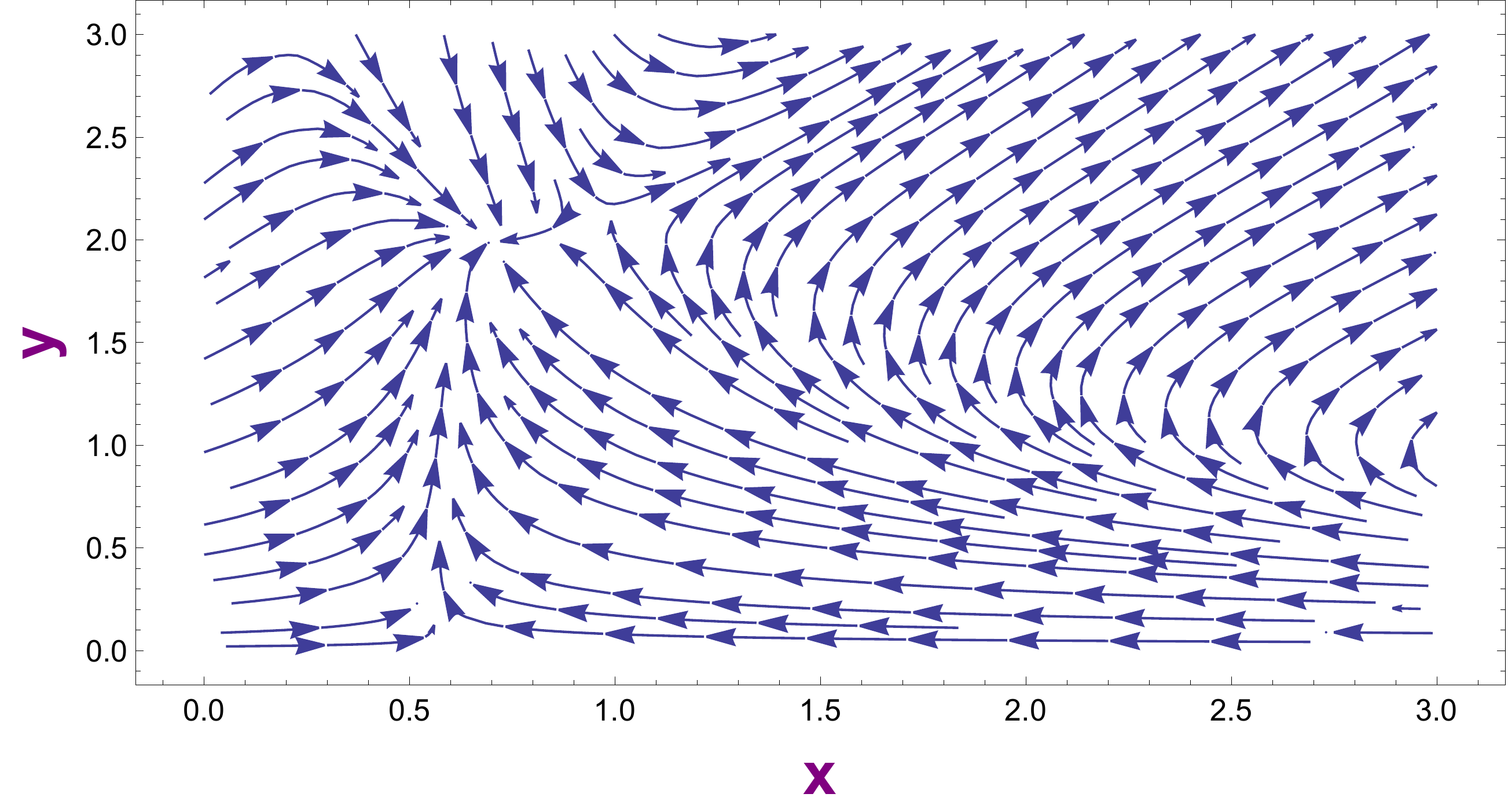}}
\caption{\label{fig:3} Parameter space(a) and phase plane(b,c) for critical point $P_3$}
\end{figure}

$P_3$: $P_3$ is the dark energy dominated solution with the universe being completely occupied by dark energy eventually. For the stability analysis of this point, considering $ Det > 0$, $Tr < 0$, and considering the four restrictions mentioned above, we finally get the feasible parameter region graph satisfying these conditions, as shown in Fig \ref{fig:3}. From the trajectory of the trace, we can see that this is a stable point, and all trajectories coming from any direction will be attracted by the point $P_3$.

\begin{figure}[h]
\centering
\subfigure{\includegraphics[width=0.5\textwidth]{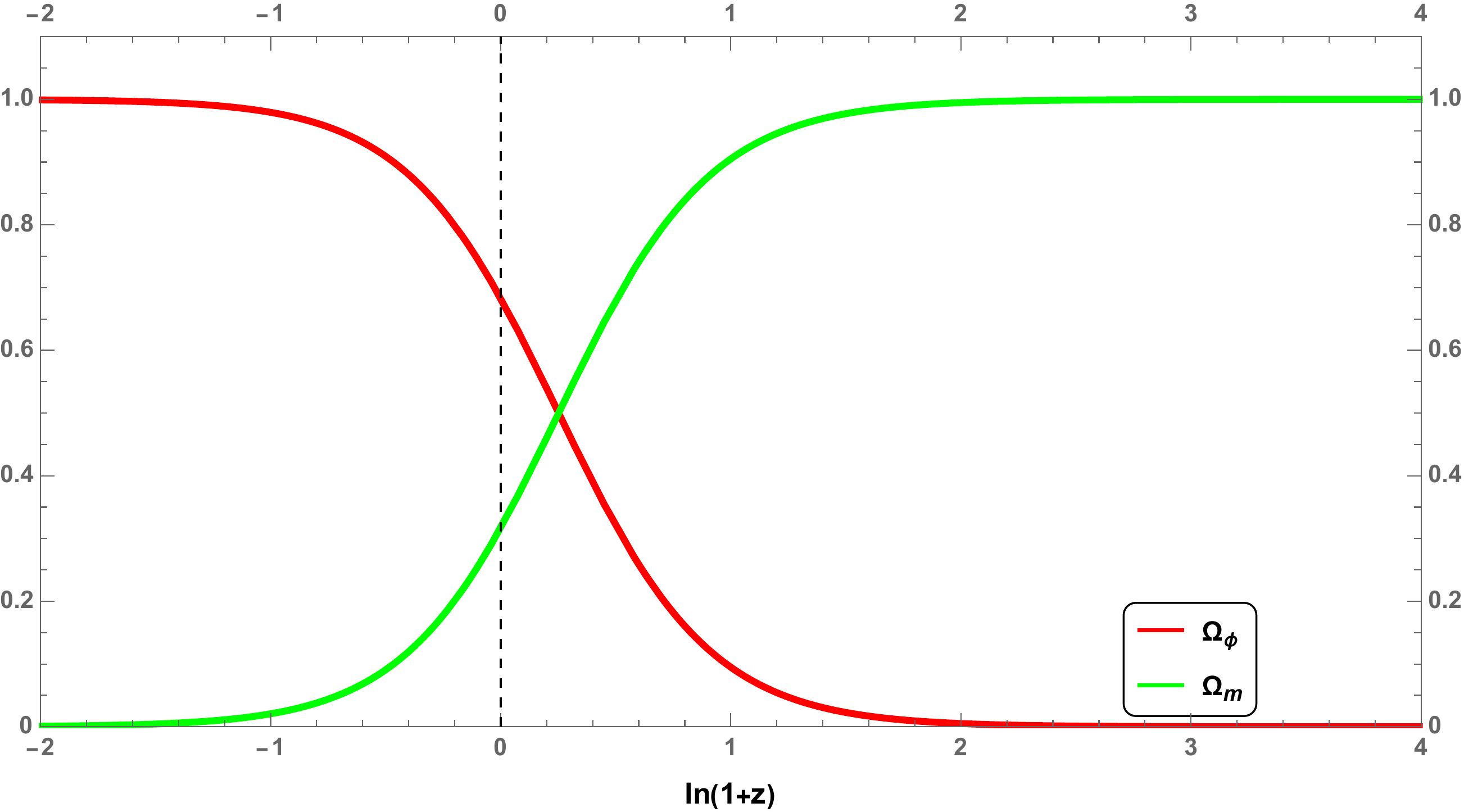}}
\subfigure{\includegraphics[width=0.5\textwidth]{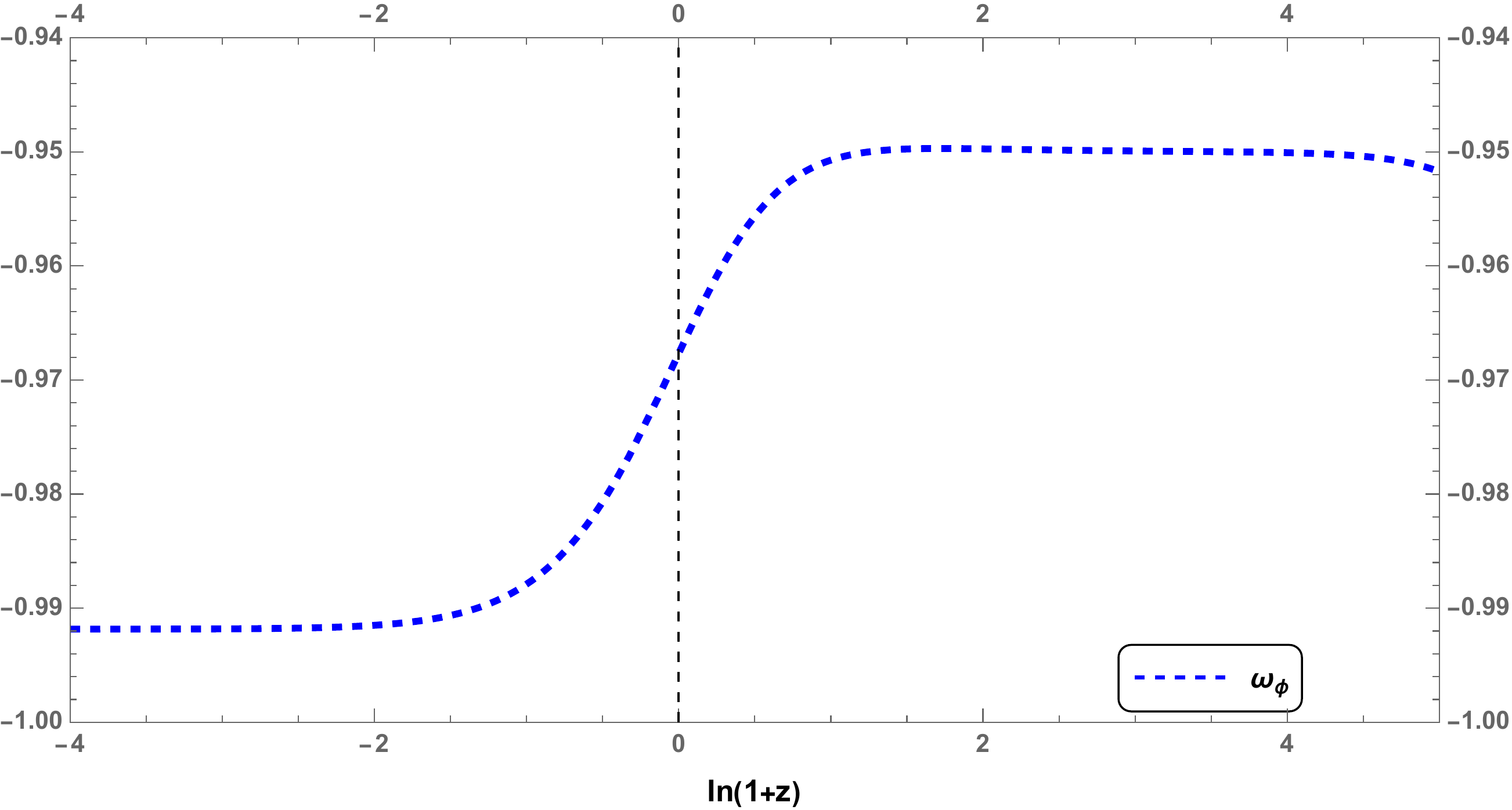}}
\caption{\label{fig:4} $P_3$: The evolution of $\Omega_{\phi}$, $\Omega_{m}$ and $w_{\phi}$ for $\left\{\alpha \rightarrow \frac{1}{20}, \lambda \rightarrow \frac{1}{100} \right\}$ with the initial conditions $x = 0.5$ and $y = 0.000078$ when $ln (1+z) = -7$.}
\end{figure}

The evolution trajectory of this point is attractive, as shown in Fig \ref{fig:4}. We choose $\alpha$ as 0.05, the initial conditions $x = 0.5$ and $y = 0.000078$ when $ ln (1+z) = 7$, and we find that $w_{\phi}$ is very close to $-0.95$ in the early stage, when the $ ln (1+z) $ is about 1, it gradually begins to decrease. In the future, $w_{\phi}$ is very close to $-0.99$. It is always greater than $- 1$ in the process of evolution. we find that the ultimate destination of the universe is that dark energy occupies the universe completely, and the final $w_{\phi}$ is $-0.99$, which accelerates the expansion of the universe. When $ ln (1+z) = 7$ evolves close to 1.5, the proportion of dark matter will gradually decrease and transform into dark energy. dark energy proportion will gradually increase, and  dominate the universe finally. Dark matter become significant more later than $P_1$.

\begin{figure}[h]
\centering
\subfigure{\includegraphics[width=0.5\textwidth]{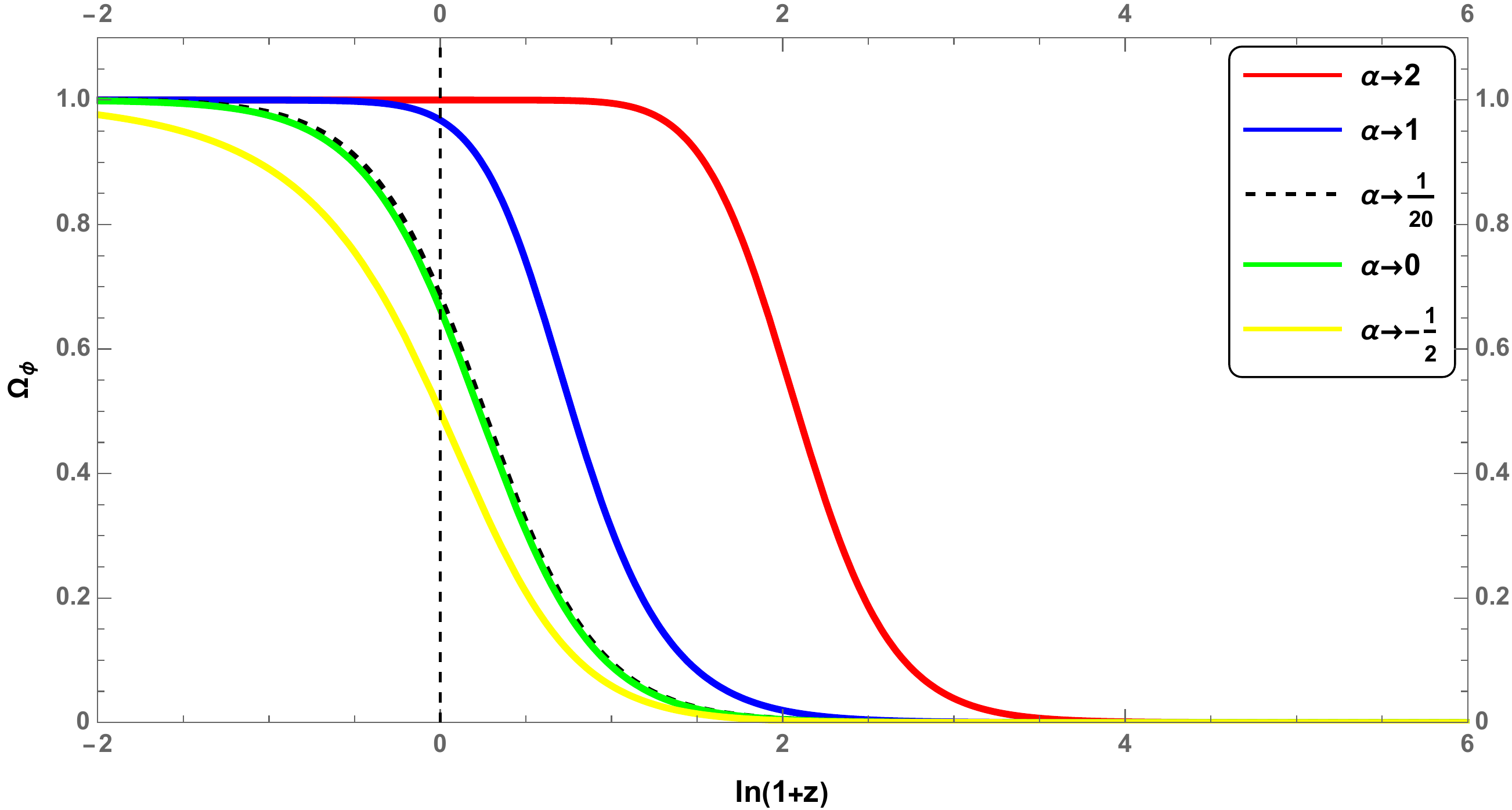}}
\subfigure{\includegraphics[width=0.5\textwidth]{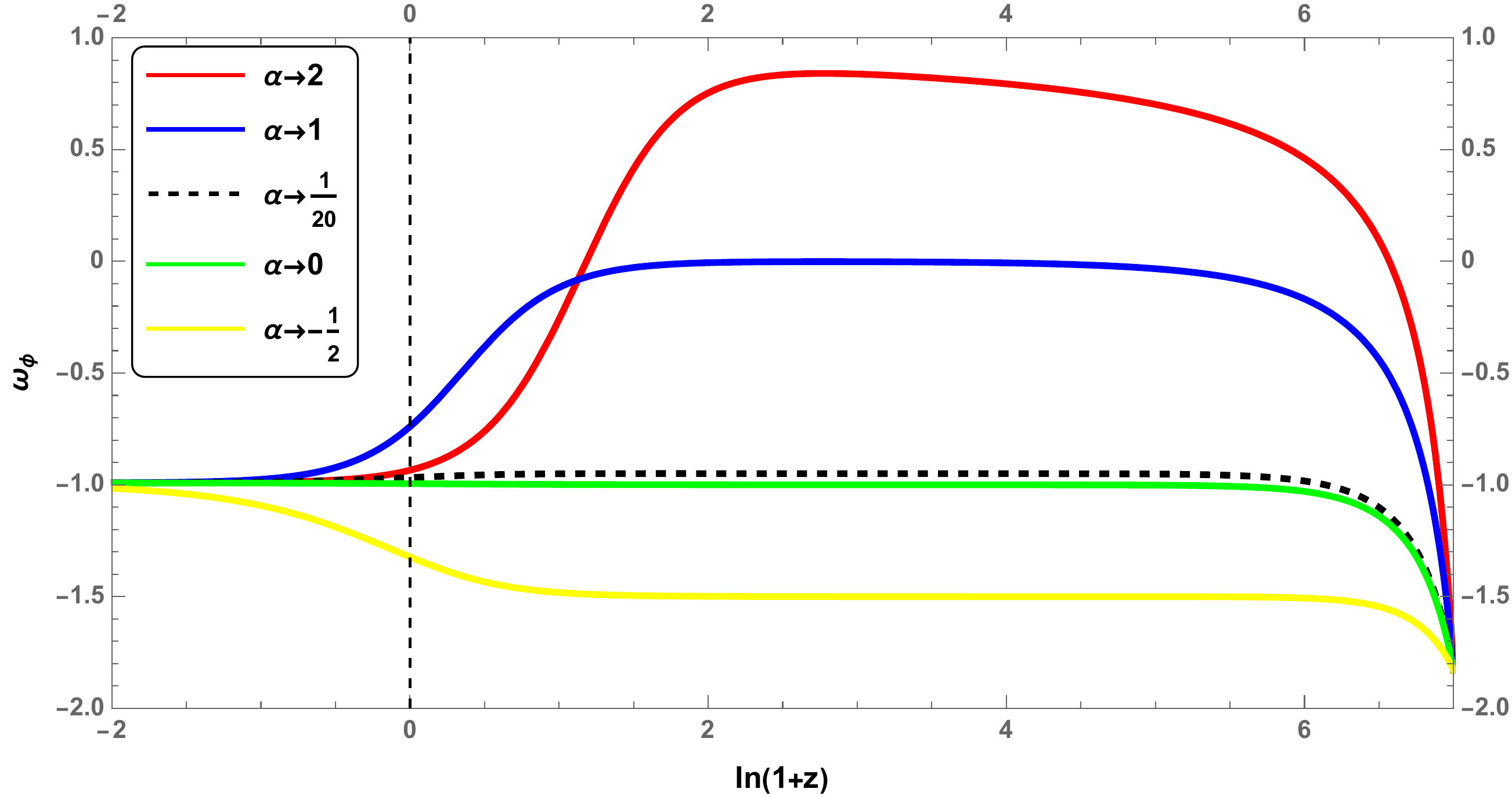}}
\caption{\label{fig:5} $P_3$: the evolution of $\Omega_{\phi}$, $w_{\phi}$ with different value of $\alpha$. The initial condition is taken as $(x, y)$=$(0.710005, 1.99184)$ around $P_3$ and $\left\{\alpha \rightarrow-\frac{1}{2},0,\frac{1}{20},1, 2; \lambda \rightarrow \frac{1}{100}\right\}$.}
\end{figure}

We find that $w_{\phi}$ and $c^2_{s}$ of $P_3$ are independent of $\alpha$ in interaction parameter $Q$ (it is necessary to emphasize that $\alpha$ can be arbitrarily taken under the condition of the stability and existence of the model satisfying our requirements). It only depends on the potential parameter $\lambda$. Although $\alpha$ does not determine the current cosmological parameters, it will determine how dark energy evolves from the early time to present. We find that when $\alpha$ is taken very small, the evolution of $\Omega_{\phi}$ is shown in Fig \ref{fig:4}, but with the value of $\alpha$ becoming larger, for example, $\alpha \rightarrow 2$, the transition from dark matter to dark energy takes place earlier. The larger the $\alpha$, the more dark energy accounts for in the earlier stage,  as shown in Fig \ref{fig:5}. For the state equation $w_{\phi}$, the value of$\alpha$ determines its evolution at the early stage when $2 < ln (1+z) <6$. We can see from Fig \ref{fig:5} that power law k-essence scalar field could behave as the stiff matter($w_{\phi}\approx 1$), the normal matter($w_{\phi}\approx 0$), the cosmological constant($w_{\phi}\approx -1$) or the phantom dark energy($w_{\phi}\approx -1.5$).

$P_4$: This point is an unstable point, so we don't consider this point.

$P_5$: Considering the four constraints mentioned above, the parameter ranges of the four conditions have no intersection which means this point does not meet the late cosmic acceleration conditions we require, so we ignore it.

\subsubsection{$x<0$}
\begin{align}
x^{\prime} &= \frac{1}{4}[-3 \alpha  x^3 y^2+2 \sqrt{3} \lambda  x^2 y+6 (\alpha -2) x-6 \sqrt{2}]\\
y^{\prime} &= \frac{1}{4} y(6+x y(3(\sqrt{2}+x) y-2 \sqrt{3} \lambda))
\end{align}
\begin{equation}
\lambda=-\frac{V_{, \phi}}{V^\frac{3}{2}}=Const
\end{equation}

In fact, the case of $x < 0$ and the case of $x > 0$ are symmetrical about $ x $, which can be seen from the Lagrangian, which means that the scalar field $\dot{\phi}$ we consider can decrease with time $t$.

\begin{figure}[h]
\centering
\subfigure[The value ranges for parameters $\lambda$ and $\alpha$ to make the critical point $P_6$ exist and stable, which is also constrained by other cosmological quantities, when $x <0$.]{\includegraphics[width=0.5\textwidth]{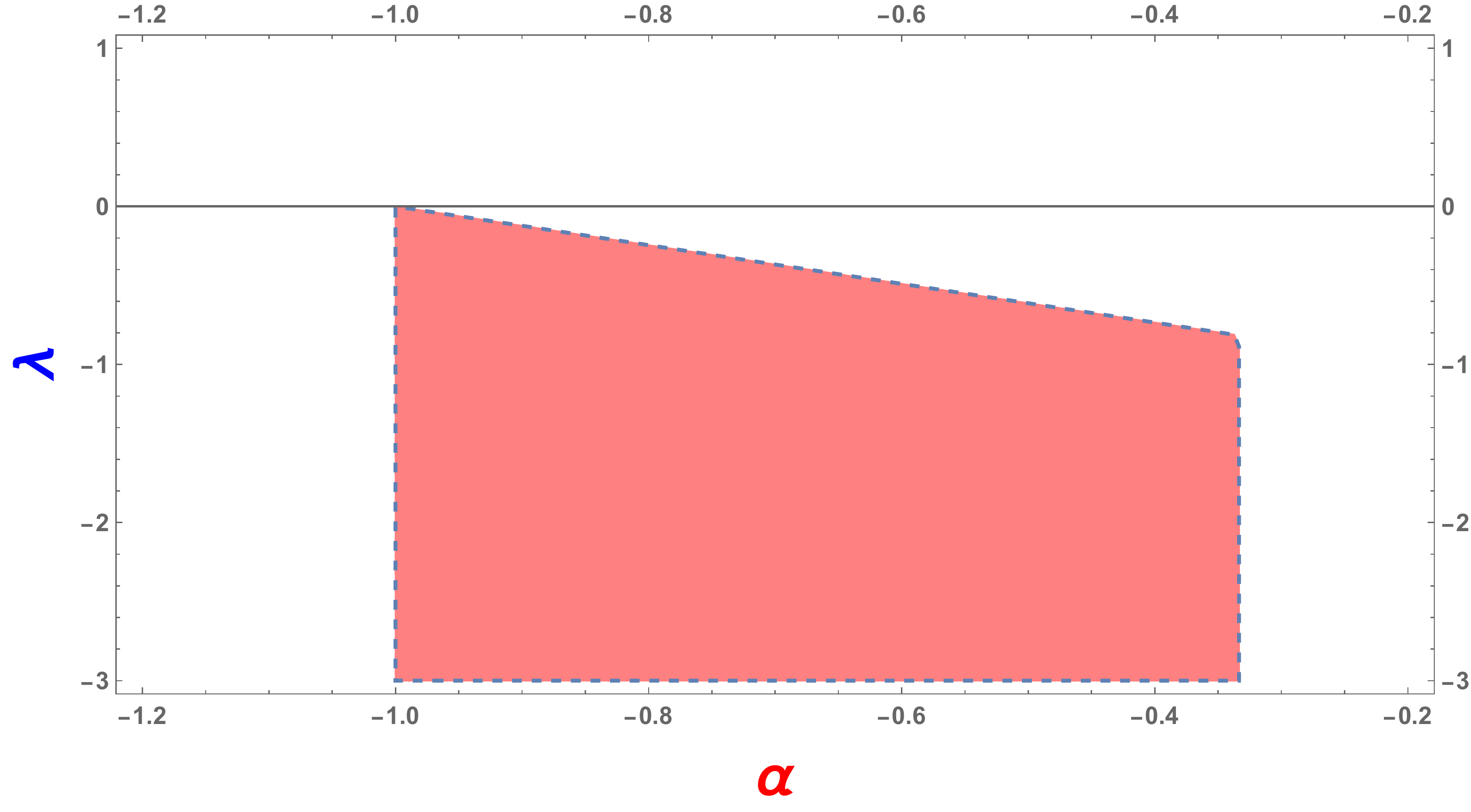}}

\subfigure[The phase plane for $\left\{\alpha \rightarrow-\frac{99}{100}, \lambda \rightarrow -\frac{12}{25}\right\}$ around the attractor $P_6$ = $(-0.71066, 1.64455)$ when $x < 0$. the red point is the final stable point.]{\includegraphics[width=0.5\textwidth]{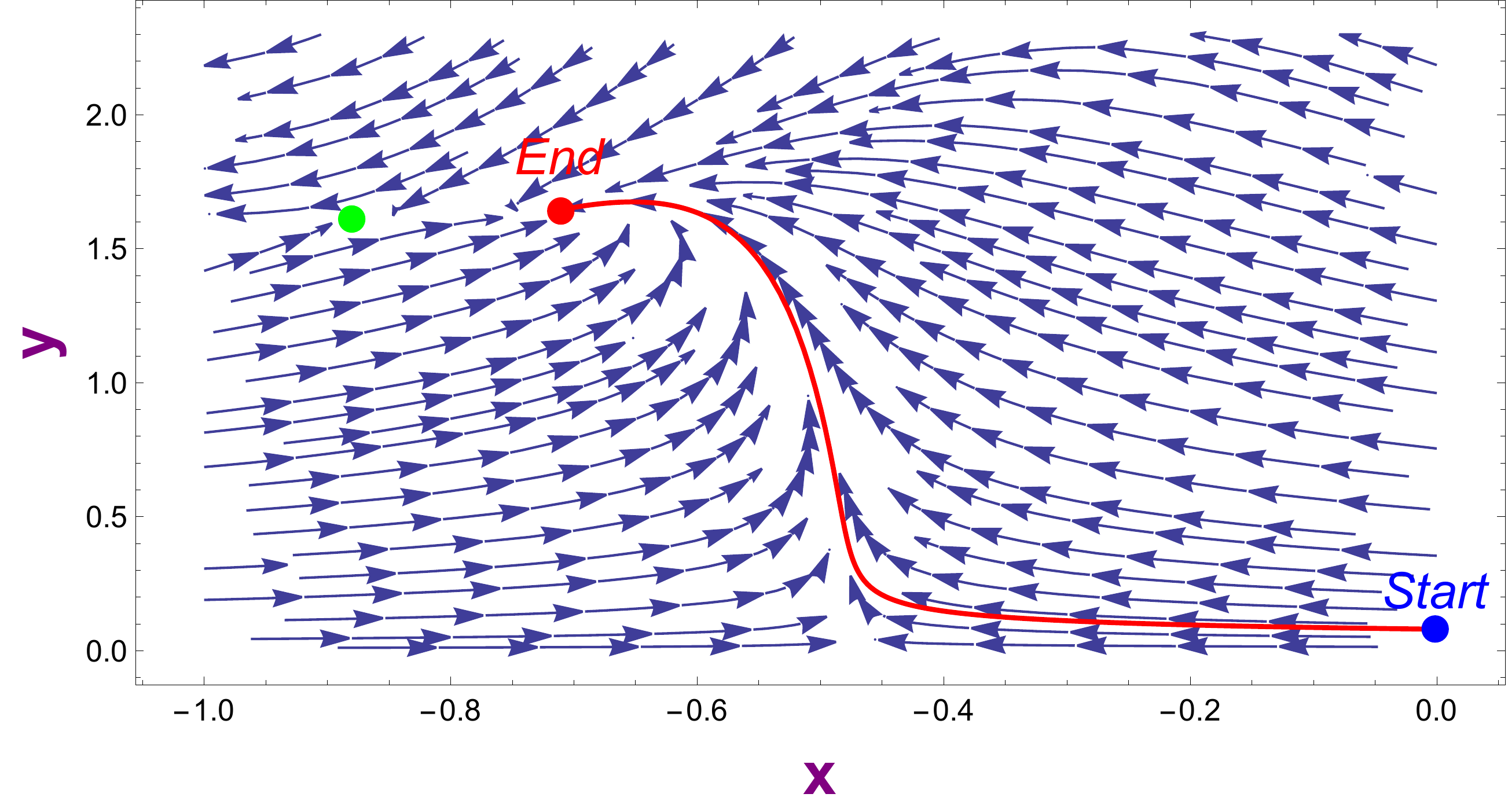}}
\caption{\label{fig:6} Parameter space(a) and phase plane(b) for critical point $P_6$.}
\end{figure}

$P_6$: We find that the evolution trajectory of cosmic parameters corresponding to this point is the same as that of point $P_1$, and the phase plane is symmetric as shown in Fig \ref{fig:1} and Fig \ref{fig:6}.

We calculate that the coordinates of the attractor $P_1$ are $(-0.71066, 1.84262)$, and the corresponding $\Omega_{\phi}$ and $w_{\phi}$ are 0.682946 and -0.99. The result given by $P_6$ is exactly the same as that given by $P_1$, except that $x$ is symmetrical due to $ \dot{\phi} >0 $ or $ \dot{\phi} <0 $, which does not change the result.

$P_7$ , $P_8$ and $P_{10}$ are the same as $P_2$ , $P_4$ and $P_{5}$ in the case of $x > 0$ respectively, so we would not consider them here.

\begin{figure}[h]
\centering
\subfigure[The value ranges for parameters $\lambda$ and $\alpha$ to make the critical point $P_9$ exist and stable, which is also constrained by other cosmological quantities, when $x <0$.]{\includegraphics[width=0.5\textwidth]{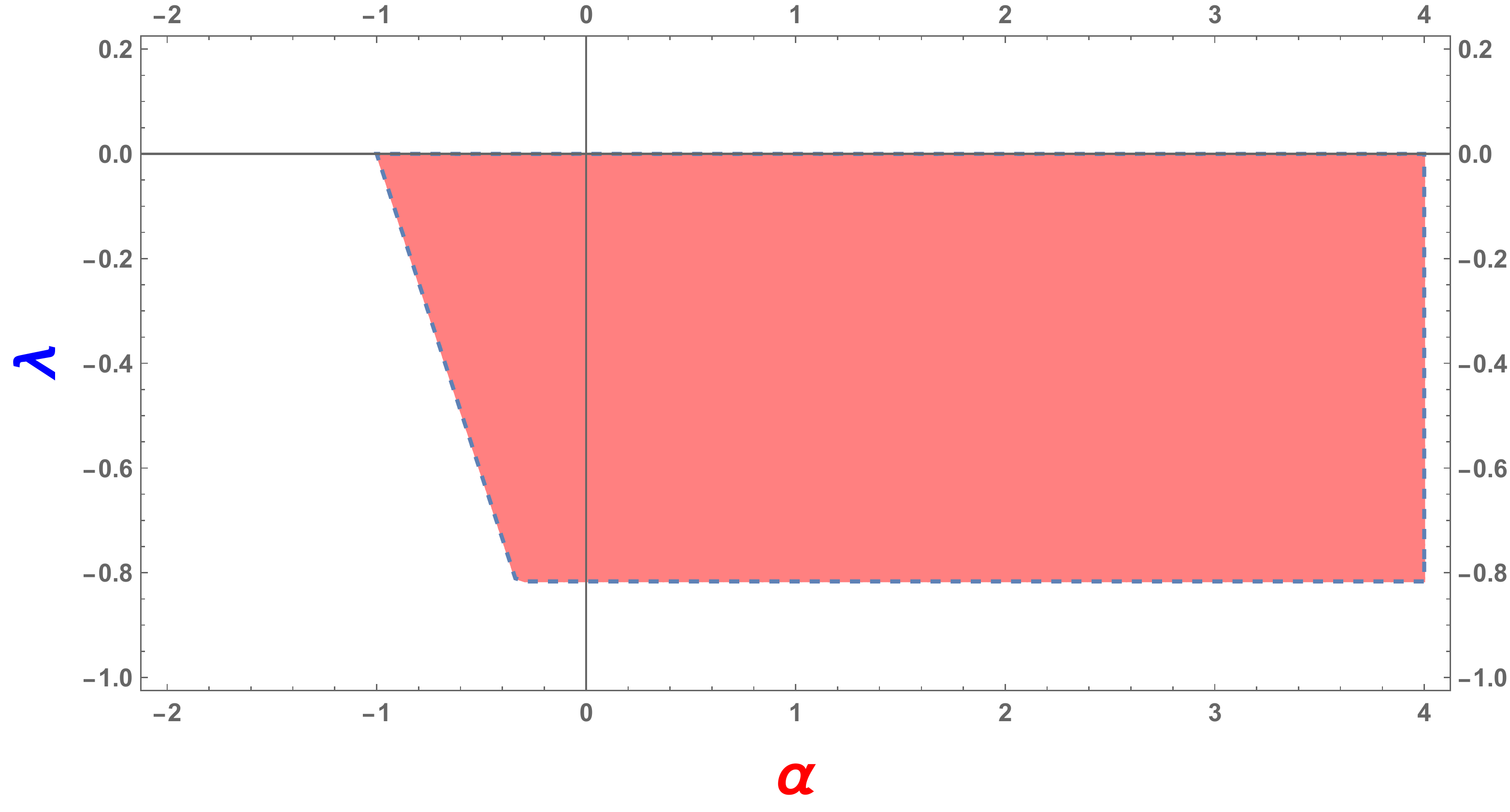}}
\subfigure[The phase plane for $\left\{\alpha \rightarrow\frac{1}{20}, \lambda \rightarrow -\frac{1}{100}\right\}$ around the attractor $P_9$ = $(-0.710005, 1.99184)$ when $x < 0$.]{\includegraphics[width=0.5\textwidth]{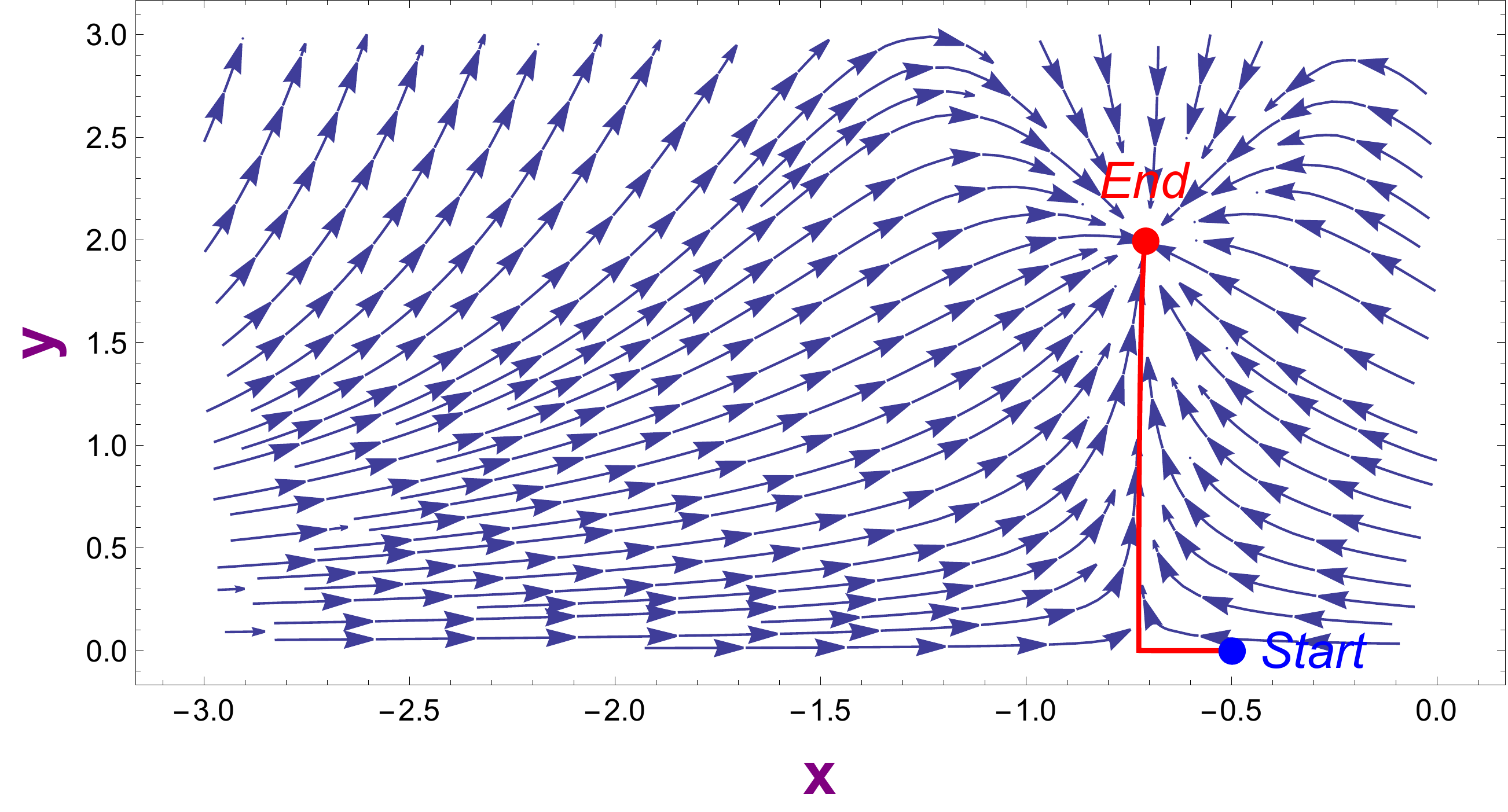}}
\subfigure[The phase plane for $\left\{\alpha \rightarrow-\frac{1}{2}, \lambda \rightarrow -\frac{1}{100}\right\}$ around the attractor $P_9$ = $(-0.710005, 1.99184)$ when $x < 0$.]{\includegraphics[width=0.5\textwidth]{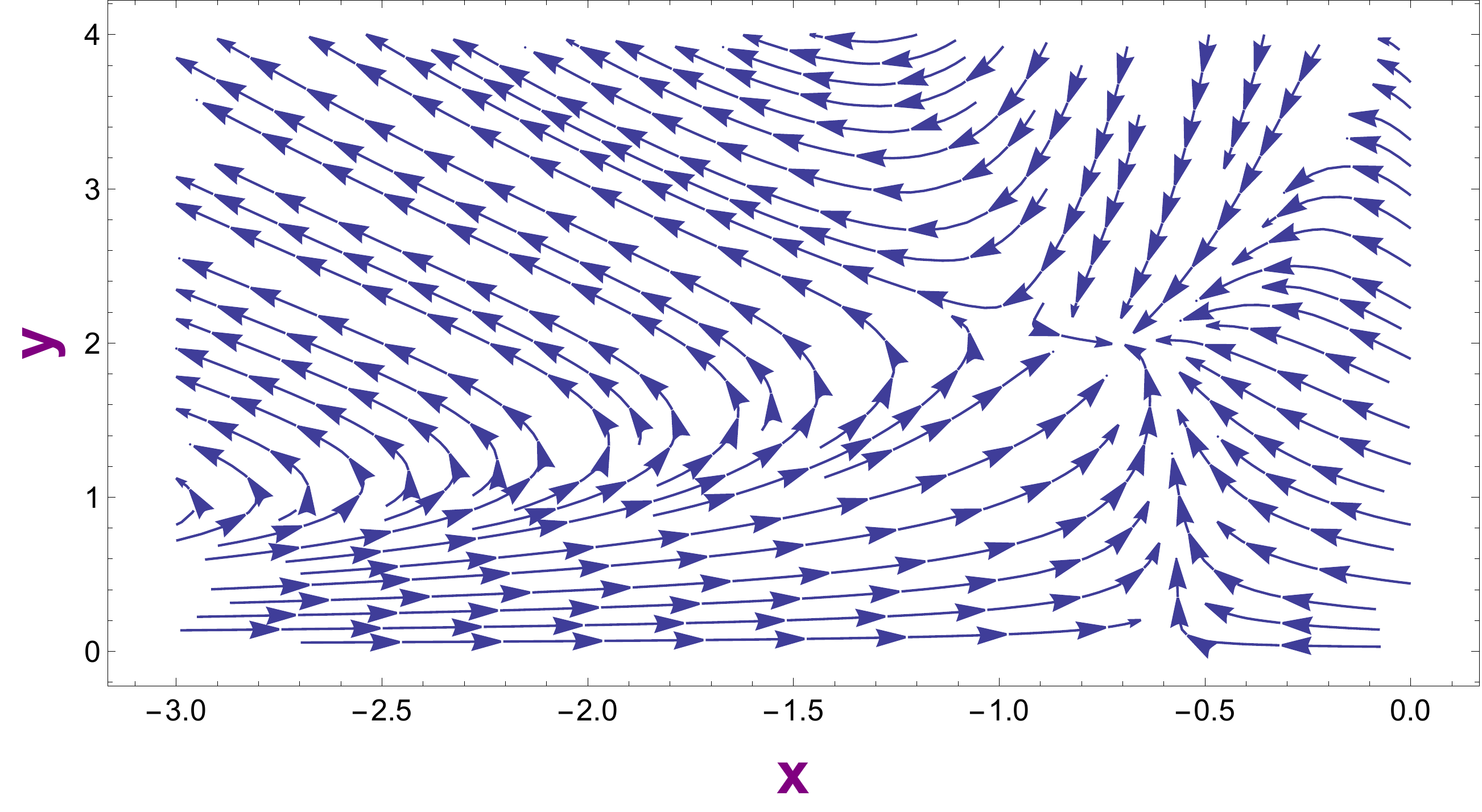}}

\caption{\label{fig:7} Parameter space(a) and phase plane(b,c) for critical point $P_9$.}
\end{figure}

$P_9$: We find that the evolution trajectory of cosmic parameters corresponding to this point is the same as that of point $P_3$, and the phase plane is similar symmetric as shown in Fig \ref{fig:3} and Fig \ref{fig:7}.

\section{Conclusions}

We investigated the dynamical system of the power law k-essence scalar field  $F(X)= -\sqrt{X} + X$ with a new interaction $Q = \alpha\rho _m\rho _{\phi }H^{-1}$. Among all the ten critical points, there are two kinds of meaningful solutions. one is Scaling-like DE solution and the other is DE dominated solution.

\begin{figure}[H]
\centering
\includegraphics[width=0.5\textwidth]{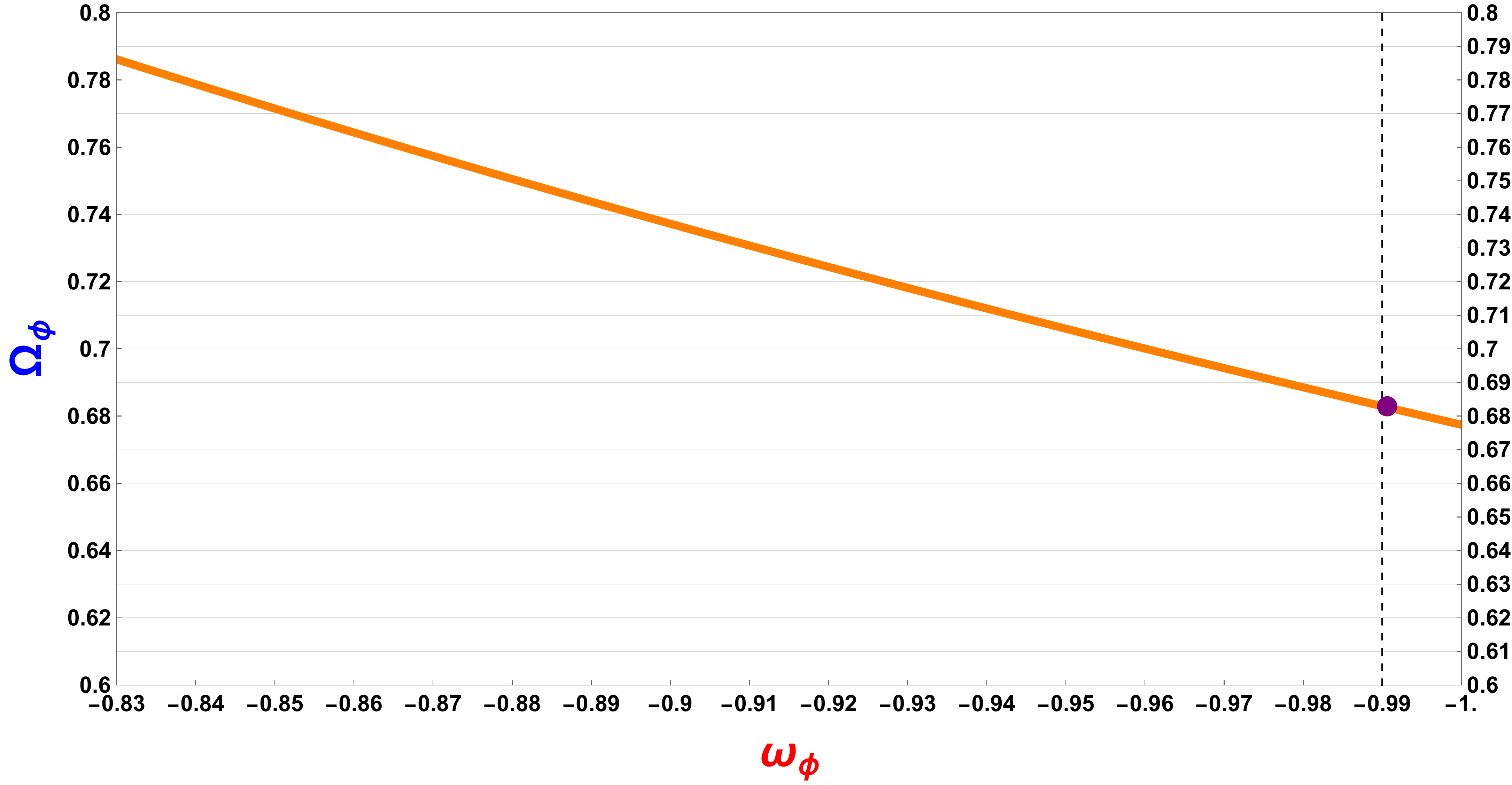}
\caption{\label{fig:8} the relationship between $\Omega_{\phi}$ and $w_{\phi}$ of the final universe evolution under the Scaling-like DE solution, and the vertical line is the value of $P_1$ which $\Omega_{\phi}$ is 0.682767 and $w_{\phi}$ is -0.99065 in this paper.}
\end{figure}

For $P_1$, this is a new Scaling-like DE solution. Compared to the same scalar field model $\mathcal{L}(\phi, X)=(-\sqrt{X} + X)V(\phi)$ with no interaction between dark energy and matter\cite{Yang_2014}, there is no Scaling-like DE solution.  $Scaling$ solution generally means that the energy of scalar field and matter coexist, so neither dark energy nor dark matter will dominate our universe separately. Moreover, for the $Scaling$ solution in previous studies, the energy of scalar field tracks the background material and behaves as dark matter$(w_{\phi}=0)$. So it is impossible to accelerate the expansion of the universe for previous $Scaling$ solution. But for the Scaling-like DE solution here, our universe will have accelerated expansion since $w_{\phi} < -\frac{1}{3}$. For our Scaling-like DE solution, the universe will be dominated by both dark energy and matter. According to and the current status of our universe could be the final  As shown in Fig \ref{fig:8}, here we plot the relationship between $\Omega_{\phi}$ and $w_{\phi}$ under the Scaling-like DE solution.

$P_3$ is another important point with cosmological meaning. The universe will be dominated by dark energy, and the value of cosmological parameters is only related to the potential energy parameter $\lambda$. However, $\alpha$ determines how the dark energy and matter evolved from early time to today. When $\alpha$ is very large, for example: $\alpha = 2$, we find that dark matter began to transform towards dark energy very early, EoS of dark energy has increasing until recent, and then decrease to $-1$ recently, subsequently  dark energy began to accelerate cosmic expansion.

Due to the limitation of sound velocity $c_{\mathrm{s}}^{2}$, we find that for the Scaling-like DE solutions $P_1$, $\alpha$ must satisfy the condition $-1<\alpha<1 $, so this restriction condition determines that the 
lower limit of $w_{\phi}$ corresponding to $P_1$ is about $-0.99$, which is still in the framework of Quintessence. Similarly, for $P_3$, $\lambda$ must be greater than 0 and less than 6. Therefore, the lower limit of $w_{\phi}$ corresponding to these two points is about $-0.99$, which is still under the framework of Quintessence. Therefore, for the new interaction model we considered, the EoS of dark energy can not be less than or equal to $-1$, so this model will not lead to the final Big Rip\cite{article1}, even though it could be less than $-1$ in the past.

 \bibliography{Reference}

\end{document}